\documentstyle[12pt]{article}
\begin{document}

\noindent P.~N.~Lebedev Institute Preprint     \hfill
                                        FIAN/TD/ 06-92\\
I.~E.~Tamm Theory Department       \hfill
\begin{flushright}{May 1992}\end{flushright}
\begin{center}
\vspace{0.1in}{\Large\bf  GAUGE FIELDS CONDENSATION AT FINITE TEMPERATURE}
\\
\vspace{0.3in}
{\large  O.~K.~Kalashnikov}\\
\medskip  {\it Department of Theoretical Physics} \\ {\it  P.~N.~Lebedev
Physical Institute} \\ {\it Leninsky prospect, 53, 117 924, Moscow,
Russia}\footnote{E-mail address: theordep@sci.fian.msk.su}\\

\end{center}

\vspace{1.5cm}

\centerline{\bf ABSTRACT}

\begin{quotation}

The two-loop effective action for the SU(3) gauge model in a constant
background field ${\bar A}_0(x,t)=B_0^3T_3+B_0^8T_8$ is recalculated
for a gauge with an arbitrary
$\xi$-parameter. The gauge-invariant thermodynamical potential is found and
its extremum points are investigated. Within a two-loop order we find that
the stable nontrivial vacuum is completely equivalent to the trivial one
but when the high order corrections being taken into account the indifferent
equilibrium seems to be broken. Briefly we also discuss the infrared
peculiarities and their status for the gauge models with a nonzero
condensate.
\end{quotation}

\newpage
At present the great attention is paid to studying the W-condensation of
the gauge fields at a finite temperature. The properties of a new vacuum
state which arises from the gluon condensation have many interesting
peculiarities and a number of models with finite densities of external
sources have been intensively studied by many authors (at first at zero
temperature [1] and then at a finite one [2,3,4]). But besides the
induced condensation (which was studied in the above mentioned papers)
a nonzero condensate of gluomagnetic fields can occur even without any
external sources (as it was discussed in [5]) and this phenomenon is
necessary to take into account in solving many problems of a non-Abelian
gauge theory. Nevertheless, a possibility for the gauge fields condensate to
arise spontaneously in the SU(N) gauge models is not proved yet although this
problem has been recently studied in many papers [6,7,8]. Simultaneously the
status of the infrared (IR) peculiarities for the modified theory is also
discussed and many authors believe that IR properties seem to change when the
nontrivial vacuum takes place. However, the gauge-invariant
thermodynamical potential was not found in papers [6,7] and
therefore a number of predictions (in particular, the possibility of
such condensation) should be verified.

Below the SU(3) gauge model in a constant background field
${\bar A}_0(x,t)=B_0^3T_3+B_0^8T_8$ is
studied. Here we use some results of paper [7] (and sometimes
those of paper [8] for SU(2) model) to find the gauge-invariant
thermodynamical potential and its extremum points. However
if only a two-loop effective action is considered
all the found minimums (including a trivial one) are equivalent and
the high order corrections should be taken into account to build
a nontrivial vacuum. A good candidate for such vacuum is presented within
SU(2) model but the problem still remains until the multi-loop effective
action for this model is calculated. These multi-loop corrections are also
important when IR peculiarities and their status are determined for the
theory with a nonzero condensate. Although within a two-loop order
we find that all essential IR peculiarities reproduce themselves it does not
exclude a possibility that the exact IR scenario will be completely different
from the standard one with a trivial vacuum.

Our formalism is built by
using the standard Green function technique at $T\neq 0$ in the background
gauge with an arbitrary parameter $\xi$. The effective action $W({\bar A}_0)$
is defined as a functional integral in periodic fields
\begin{eqnarray}
&& \exp[-W({\bar A}_0)V/T]=\\
&& =N\int{\cal D}Q{\cal D}C{\cal D}{\bar C}
\exp{\left\{-\int\limits_0^{\beta}d\tau \int d^3x
({\cal L}_A+{\cal L}_{g.f.}+{\cal L}_{g.h.}-Q\cdot J)\right\}}\nonumber
\end{eqnarray}
where N is a temperature independent normalization factor (here
$\beta=1/T$), $V$ is a space volume; $Q_{\mu}$, $C$ and ${\bar C}$ are the

quantum gauge fields and ghosts, respectively; and J is an external source.
The gauge field Lagrangian has the usual form
$$
{\cal L}_A+{\cal L}_{g.f.}=\frac{1}{4}(G_{\mu\nu}^a)^2+\frac{1}{2\xi}
[({\bar {\cal D}}_{\mu}A_{\mu})^a]^2  ,
\eqno{(2)}
$$
but the gauge fields $A_{\mu}^a$ are decomposed in the quantum part
$Q_{\mu}^a$ and the classical constant one ${\bar A}_{\mu}^a$ (here
$A_{\mu}^a=Q_{\mu}^a+{\bar A}_{\mu}^a$). In Eq.(2) the covariant derivative
is ${\bar {\cal D}}_{\mu}^{ab}=\partial_{\mu}{\delta}^{ab}+gf^{acb}{\bar
A}_{\mu}^c$ and the gauge field strength tensor is determined as follows
$G_{\mu\nu}^a=({\bar{\cal D}}_{\mu}Q_{\nu})^a-({\bar{\cal
D}}_{\nu}Q_{\mu})^a+gf^{abc}Q_{\mu}^bQ_{\nu}^c$.
The ghost Lagrangian is built according to the standard rules
$$
{\cal L}_{g.h.}=-{\bar C}{\bar {\cal D}}_{\mu}{\cal D}_{\mu}C
\eqno{(3)}
$$
where all ghost fields although have the Fermi commutation relations but at
$T\neq 0$ they are quantized as Bose fields with the aid of the even
frequencies ($\omega_n=2\pi n/\beta$) [9]. The essential advantage of the
background gauge is that only one Z-factor is necessary for its
renormalization and many exact results are proved much easier within this
technique. However,the essential attribute of this gauge is a special
prescription of using background fields and for many cases to build the
correct sequence of calculations (which provides a gauge invariance
explicitly) is a rather nontrivial task. Today this prescription is known
only for $T=0$ case (see e.g. [10] and other papers within it) but
for the typical $T\neq 0$ perturbative calculations the appropriate ansatz
is not evident and gauge invariance needs checking
step by step. To solve this problem (when the thermodynamical potential is
calculated perturbatively) a rather simple ansatz has been recently used
(at first in [11] and then in [4] for the non-Abelian case) and being
model-independent this prescription presents a reliable way of treating
the different gauge models. Here this scenario is
reproduced to determine the two-loop thermodynamical potential after the
effective action found in [7] has been recalculated for
eliminating the noticed misprints.

The perturbative graphs (despite the nonzero external fields)
have the standard form where
instead of the background fields it is convenient to use the following
scalar variables
$$
x=\frac{gB_0^3}{{\pi}T}\,,\qquad y=\frac{gB_0^8}{{\pi}T}
\eqno{(4)}
$$
Each time the background fields are taken into account exactly and after all
calculations being performed these fields should be eliminated
selfconsistently to obtain the gauge-invariant result. Our calculations
follow those from paper [7] which should be repeated since for the
gauge with an arbitrary $\xi$-parameter these results are checked nowhere.

The one-loop effective action for the SU(3)-model has the standard form
$$
W^{(1)}(x,y)/T^4=\frac{4{\pi}^2}{3}\left\{B_4(0)+B_4(\frac{x}{2})+
B_4[{\frac{1}{4}}(x+{\sqrt{3}}y)]+B_4[{\frac{1}{4}}(x-{\sqrt{3}}y)]\right\}
\eqno{(5)}
$$
and represents the independent contributions of eight gauge
fields after some kind of diagonalization. The two-loop action is more
complicated and after our corrections being made it is found as follows
\setcounter{equation}{5}
\begin{eqnarray}
&& W^{(2)}(x,y)/T^4=\nonumber\\
&=& \frac{g^2}{2}\left\{B_2^2(\frac{x}{2})+B_2^2(\frac{v}{2})
+B_2^2(\frac{u}{2})+2B_2(0)[B_2(\frac{x}{2})+B_2(\frac{v}{2})
+B_2(\frac{u}{2})]+\right.\nonumber\\
&+& \left.B_2(\frac{v}{2})B_2(\frac{u}{2})+B_2(\frac{x}{2})[B_2(\frac{v}{2})+
B_2(\frac{u}{2})]\right\}+\\
&+& \frac{g^2}{3}(1-\xi)\left\{2B_1(\frac{x}{2})[B_3(\frac{x}{2})+
\frac{1}{2}(B_3(\frac{v}{2})+B_3(\frac{u}{2}))]+2B_1(\frac{v}{2})
[B_3(\frac{v}{2})+\right.\nonumber\\
&+& \left.\frac{1}{2}(B_3(\frac{x}{2})-B_3(\frac{u}{2}))]+
2B_1(\frac{u}{2})[B_3(\frac{u}{2})+\frac{1}{2}(B_3(\frac{x}{2})-
B_3(\frac{v}{2}))]\right\}\nonumber
\end{eqnarray}
where some signs and coefficients of the terms depended on a gauge parameter
$\xi$ are different from the paper [7].
Here we omitted our calculations (their
details one can find in papers [6,7]) since they have not any new elements
besides the pointed above changes which however are very essential for what
follows. In the formulae (5) and (6) $B_n(z)$ are the standard Bernoulli
polynomials and the new variables are $v=\frac{1}{2}(x+\sqrt{3}y)$ and
$u=\frac{1}{2}(x-\sqrt{3}y)$.

Our next result is to present (and then to use) ansatz (earlier checked in
papers [4,11]) that gives an opportunity to build step by step within
perturbative calculations the gauge-independent thermodynamical potential
if the effective action is known. For building this potential within a
chosen approximation (i.e. for minimizing the known effective action) we
must exploit the specially found extremum equations which result from
the effective action calculated on
the previous stage (but not the same one) since the location of
extremum points is a gauge-dependent quantity itself.
Here this means that for calculating the thermodynamical potential of
order $g^2$ one should use the one-loop effective action for defining the
necessary extremum points and then substitute
the solutions obtained within these equations
in $W^{(2)}(x,y)$. Using Eq.(5) these equations are found to be
\begin{eqnarray} && B_3(\frac{\bar x}{2})+\frac{1}{2}
[B_3(\frac{\bar v}{2})+B_3(\frac{\bar u}{2})]=0\nonumber\\ && B_3(\frac{\bar
v}{2})-B_3(\frac{\bar u}{2})=0
\end{eqnarray}
and after they being used all
terms in Eq.(6) which are proportional to $\xi$
cancel each other. We believe that such cancelation will take place step by
step although each time the extremum points will change. Only when these
points are substituted in Eqs.(5) and (6) the arisen expression has a
physical meaning and presents the gauge-independent thermodynamical potential
of
the model studied.

Within Eq.(7) there are six different solutions (those with
$\bar y={\pm}{\bar x}/\sqrt{3}$ are the same) but only two  are not trivial
$$
\bar y={\pm}{\bar x}/\sqrt{3}\,,\qquad \bar x=1,2
\eqno{(8)}
$$
For these points the thermodynamical potential has the form
$$
\Omega/T^4=\frac{8{\pi}^2}{3}[B_4(0)+B_4(\frac{\bar x}{2})]+
\frac{3g^2}{2}[B_2(0)+B_2(\frac{\bar x}{2})]^2
\eqno{(9)}
$$
and one can check that its minimum (if $ g^2<{\pi}^2$  )
takes place only when $x=2$
$$
\Omega/T^4=-{\frac{8{\pi}^2}{3}}{\frac{1}{15}}+{\frac{3g^2}{2}}(\frac{1}{3})^2
\eqno{(10)}
$$
Unfortunately the found potential is the same as one for
the trivial case $x,y=0$ and this fact means that the scenario with
$x,y\neq 0$ is not reliable. For $ g^2>2{\pi}^2$ another nontrivial
solution (with $y={\pm}x/\sqrt{3}$ and $x=1$ ) is preferable but
for this case the two-loop thermodynamical potential should be recalculated
since the high orders
corrections can be essential. Moreover within a two-loop approximation (see
Eqs.(5),(6) and (7) ) all solutions with $x=0 ,(\sqrt{3}/2)/y=1,2$ are
completely identical those with
$ y={\pm}x/\sqrt{3}, x=1,2 $ and this fact also decreases
our chance of building the real picture within calculations made.

 When the high order corrections being taken into account the indifferent
equilibrium seems to be broken but for solving this question finally
the exact three-loop effective action (that is better than the
nonperturbative one) should be calculated. Of course, these
calculations are very complicated and they are likely to be possible only
for the constant external field ,for example, within SU(2) model [8]
where any algebra is easily performed and the found effective action
(including the two-loop graphs) has a rather simple form

\setcounter{equation}{10}
\begin{eqnarray}
W(x)/T^4&=&
\frac{2}{3}\pi^2[B_4(0)+2B_4(\frac{x}{2})]+
\frac{g^2}{2}[B_2^2(\frac{x}{2})+2B_2(\frac{x}{2})B_2(0)]\nonumber\\
&+&\frac{2}{3}g^2(1-\xi)B_3(\frac{x}{2})B_1(\frac{x}{2})
\end{eqnarray}

The three-loop effective action $W^{(3)}(x)$ will undoubtly display a very
nontrivial dependency from the gauge parameter $\xi$ but this action should
be considered together with its own (in this case two-loop) extremum
equations which are necessary to cancel all the $\xi$ -
dependent terms and to transform the studied action into the gauge-invariant
quantity. The equation for determining the new extremum points is
found within Eq.(11) and we can rewrite it as follows
\begin{eqnarray}
&&B_3(\frac{{\bar x}}{2})+ \frac{3g^2}{8\pi^2}B_1(\frac{{\bar
x}}{2})[B_2(\frac{{\bar x}}{2})+ B_2(0)]\nonumber\\
&& +\frac{g^2}{8\pi^2}(1-\xi)[3B_2(\frac{{\bar x}}{2})B_1(\frac{{\bar x}}{2})+
B_3(\frac{{\bar x}}{2})]=0
\end{eqnarray}
where the $\xi$-dependent terms appear explicitly as it should be in
a more complicated (real) case. Within Eq.(12)
besides the previously known points (${\bar x}_1=0$ and
${\bar x}_1=1$) the new solution arises
\begin{equation}
{\bar x}_3=1\pm
\left\{1-
\frac{
\frac{g^2}{2\pi^2}[1+(1-\xi)/2]}
{
1+\frac{3g^2}{8\pi^2}[1+\frac{4}{3}(1-\xi)]}
\right\}^{1/2}
\end{equation}
which is a good candidate for a nontrivial vacuum although the problem remains
to prove that the solution (13) is correspond to the minimum (or better the
absolute one) of the thermodynemical potential to come.

The status of the standard infrared (IR) divergencies (e.g.see the review
[12]) and especially the main IR ones of the transversal
part of $\Pi_{\mu\nu}$ for the gauge models with a
nonzero $A_0$-condensate is not finally defined either and the additional
analysis is necessary. All approximations studied above are very trivial for
solving this question while the more complicated calculations (which could
change the situation) are not made yet. However it is known [13] that the
standard polarization tensor for SU(N) models changes its form when
the constant electric field is applied
\begin{eqnarray}
&& \Pi_{ij}({\bf p},{{\hat p}_4})=\nonumber\\ &=& [\delta_{ij}-\frac{p_i
p_j}{{\bf p}^2}]A({\bf p},{\hat p}_4) +\frac{p_i p_j}{{\bf p}^2}\frac{{\hat
p}_4^2}{{\bf p}^2} [\Pi_{44}({\bf p},{\hat p}_4)-2i{g^2}{\rho}/{\hat p}_4],
\nonumber\\ && \Pi_{i4}({\bf p},{\hat p}_4)=-\frac{p_i \hat p_4}{{\bf p}^2}
[\Pi_{44}({\bf p},{\hat p}_4)-2i{g^2}{\rho}/{\hat p}_4]
\end{eqnarray}
and this form
demonstrates that the new tensor structures are
generated by external fields (the polarization tensor is not transversal if
externel $\rho \neq 0$) and
all scalar functions only depend on the shifted momenta ${\hat p}_4=p_4+{\bar
A}_0 $. If $ {\bar A}_0 $ points are not trivial (e.g. as in Eq.(13)) there
is a chance of softening IR peculiarities since the condition ${\hat p}_4=0$
is not fulfilled when the summation on frequencies being performed. But
within a two-loop approximation all stationary solutions for ${\bar A}_0$
are proportional $2\pi T$ (as well as $p_4$) and IR divergencies remain
the same although their source is shifted (now all IR divergencies arise
when ${\hat p}_4=0 $ but not $p_4=0$ as earlier). Within multi-loop
calculations
IR status of the transversal part $\Pi_{\mu\nu}$ is not so evident especially
if a gauge invariance of
the thermodynamical potential results from the made
calculations and a found solution for $ {\bar A}_0 $ realizes the minimum
(but not simply the extremum) of the investigated potential. Moreover all the
discussed IR limits will essentially depend on the gauge fixed (the signal is
a deformation of ${\bar A}_0$-point in Eq.(13) when the $\xi$-parameter
changes) and the appropriate choice of the gauge parameter $\xi$ is necessary
to solve the problems stated. Of course, IR limit of the longitudinal part
${\Pi}_{\mu\nu}$ changes but all essential IR divergencies of any
gauge theory arise within its transversal sector where the external magnetic
field (see e.g. review [14]) rather than electric
one is a more effective instrument for eliminating IR instability.

In conclusion we repeat once more that the perturbative effective action being
put on the extremum points (that is the thermodynamical potential)
does not depend on the gauge fixing parameter only if the special ansatz is
used for determining these points which location is a gauge-dependent quantity.
At the same time the structure of the nontrivial vacuum that results from the
sponteneously arisen condensate is not defined yet and the case of the
indifferent equilibrium or metastable one is not excluded. However, the
calculations made are not enough to reply on this problem and their
prolongation is very actual.
Here we hope that the exactly found three-loop effective action
rather than the nonperturbative one (e.g.see [15]) will be more useful for
solving all questions stated since the gauge invariance within exact
calculations provides the reliable test for the results found.

\vspace {0.5cm}
\begin{center}
{\bf References}
\end{center}

1. A.D.Linde. Phys. Lett. {\bf 86B} (1979) 39; I.V.Krive. Yad. Fiz.
{\bf 31} (1980) 1259 [Sov. J. Nucl. Phys. {\bf 31} (1980) 650].

2. O.K.Kalashnikov and H.Perez Rojas. Kratkie Soobcheniya po Fiz.
{\bf 2} (1986) 23 [Sov. Phys. Lebedev Inst. Reports (1986)]. H.Perez Rojas
and O.K.Kalashnikov. Nucl. Phys. {\bf B293} (1987) 241.

3. E.Ferrer, V. de la Incera and A.E.Shabad. Phys. Lett. {\bf B185}
 (1987) 407; Nucl. Phys. {\bf B309} (1988) 120.

4. O.K.Kalashnikov and H.Perez Rojas. Phys. Rev. {\bf D40} (1989)
 1255: O.K.Kalashnikov, L.V.Razumov
and H.Perez Rojas. Phys. Rev. {\bf D42} (1990) 2363.

5. N.Weiss. Phys. Rev. {\bf D24} (1981) 475;
O.K.Kalashnikov, V.V.Klimov and E.Casado. Phys. Lett. {\bf 114B}
 (1982) 49.  R.Anishetty. J.Phys. {\bf G10} (1984) 439.

6. V.M.Belyaev and V.L.Eletsky. Z. Phys. {\bf C45} (1990) 355.

7. K.Enqvist and K.Kajantie. Z. Phys. {\bf C47} (1990) 291.

8. V.V.Skalozub. Preprint ITP-92-12E (Kiev, 1992).

9. C.Bernard. Phys. Rev. {\bf D9} (1974) 3312; I.V.Tyutin. Lebedev
Physical Inst. Preprint N39 (1975).

10. L.F.Abbott. Nucl. Phys. {\bf B185} (1981) 189.

11. O.K.Kalashnikov, V.V.Razumov. Yad. Fiz. {\bf 50} (1989) 884
[Sov. J. Nucl. Phys. (1989)].

12. O.K.Kalashnikov. Preprint BNL-46878 NUHEP-TH-91 (October,1991).

13. O.K.Kalashnikov, V.V.Razumov. Kratkie Soobcheniya po Fiz. {\bf
1} (1990) 50 [Sov. Phys. Lebedev Inst. Reports (1990)].

14. O.K.Kalashnikov . Fortschr. Phys. {\bf 32} (1984) 525.

15. V.M.Belyaev. Phys. Lett. {\bf 254B} (1991) 153.

\end{document}